\documentclass{article}

\usepackage{geometry}
\usepackage{amsmath}
\usepackage{mathabx}
\usepackage{graphicx}
\usepackage{subfigure}
\usepackage{caption}
\usepackage{subcaption}
\usepackage{mwe}
\usepackage[T1]{fontenc}

\usepackage{xcolor}
\usepackage{rotating}
\usepackage{authblk}
\usepackage{float}

\usepackage[linktoc=all,colorlinks=true,citecolor=blue,linkcolor=blue]{hyperref}

\newcommand{\subf}[2]{%
  {\small\begin{tabular}[t]{@{}c@{}}
  #1\\#2
  \end{tabular}}%
}

\title{Impenetrable Barriers in the Phase Space of a Particle Moving Around a Kerr Rotating Black Hole}

\author[1,2,3]{Francisco Gonzalez Montoya \thanks{ f.gonzalezmontoya@leeds.ac.uk, \\f.gonzalez.montoya@ciencias.unam.mx}}

\small{

\affil[1]{Faculty of Physical Sciences and Engineering, University of Leeds, Leeds, LS2 9JT, United Kingdom}

\affil[2]{Instituto de Ciencias Físicas, Universidad Nacional Aut\'onoma de M\'exico, Av. Universidad s/n, Col. Chamilpa, Cuernavaca, Morelos, CP 62210, México }

\affil[3]{Facultad de Ciencias,  Universidad Nacional Aut\'onoma de M\'exico, Av. Universidad 3000, Circuito Exterior s/n, Coyoacán, CP 04510, Ciudad Universitaria, Ciudad de M\'exico, México}

}

\date{\today}

\begin{document}

\maketitle

\begin{abstract}

We study the phase space of a particle moving in the gravitational field of a rotating black hole described by the Kerr metric from a geometrical perspective. In particular, we show the construction of a multidimensional generalization of the unstable periodic orbits, known as Normally Hyperbolic Invariant Manifold, and its stable and unstable invariant manifolds that direct the dynamics in the phase space. Those stable and unstable invariant manifolds divide the phase space and are robust under perturbations. To visualize the multidimensional invariant sets under the flow in the phase space, we use a method based on the arclength of the trajectories in phase space known as Lagrangian descriptors in the literature. 

\end{abstract}

\newpage

\section{Introduction}

The rotating black holes result from some of the most extreme events in the universe like when a massive star reaches the end of its life and implodes, collapsing on itself and forming a stellar black hole. When many stellar black holes merge they can form a supermassive black hole like the one in the centre of many galaxies like our Milky Way. The rotation of the black holes is the result of interaction between non-colinear masses during its formation \cite{chandra, wiltshire}.

The analysis of the dynamics of particles and light around rotating black holes has been important to study diverse astronomical phenomena like the motion of the bright gas around them. The observations with new telescopes combined with the progress in data analysis techniques have made it possible to visualize a black hole using the images of the gas around it for the first time in recent years \cite{akiyama2019,akiyama2022}. 

The Kerr metric describes the gravitational field in a neighbourhood of a rotating black hole \cite{kerr}.  A test particle moving around a Kerr black hole is a Liouville integrable system \cite{carter,carter1}, its Hamiltonian function has 4 independent integrals of motion:  the Hamiltonian $H$, the energy of the particle $E$,  the component of the angular momentum around the rotation axis of the black hole $L$, and the Carter integral $Q$. Using those integrals, it is possible to study in detail the dynamics of the system. Recent analytical studies of the phase space of this system have shown:  the bifurcations of the main families of periodic orbits \cite{bizyaev}, homoclinic orbits in the equatorial plane \cite{levin1,levin2}, trajectories converging to photon sphere \cite{sneppen}, the classification of some bounded trajectories using action-angle variables \cite{levin}, the classification of radial motion \cite{compere},  null and time like geodesics \cite{chen} , the spherical orbits \cite{teo,teo1}, and other bounded trajectories relevant for astrophysics \cite{rana,rana1}.  

In this work, we take a geometric approach to study and visualize the phase space of the system. In particular, we consider invariant objects under the dynamics in the phase space that generalize the unstable hyperbolic periodic orbits and their stable and unstable invariant manifolds. Considering the global phase space, we can construct impenetrable barriers to understand the dynamics of the system.

The rotation of the black hole is characterized by a parameter $a \in [0,1]$ proportional to the angular momentum of the black hole. For $a=0$, the black hole is not rotating and its gravitational field is described by the Schwarzschild metric \cite{wald}. In this case, the spatial part of the metric is spherically symmetric, so it is natural to construct a radial effective potential energy parameterized by the angular momentum $L$ to describe the radial motion of the particle. Associated with the maximum value of this radial effective potential energy, there is a family of unstable hyperbolic periodic orbits. Their stable and unstable manifolds of unstable hyperbolic periodic orbits play a fundamental role in the analysis of the phase space of the system. In particular, for the construction of the impenetrable barriers in the phase space in the non-integrable perturbed case.

Considering those geometrical properties of the phase space of the system, we construct a Normally Hyperbolic Invariant Manifold (NHIM) taking the union of all unstable periodic orbits associated with the maximum of the radial effective potential energy for all the different directions on the sphere. The stable and unstable manifolds of this NHIM have codimension one relative to the constant energy manifold.  Therefore,  the stable and unstable manifolds of this NHIM are impenetrable barriers that direct the dynamics in the phase space.

An important property of the NHIMs and their stable and unstable invariant manifolds is that they are robust under generic perturbation of the Hamiltonian \cite{fenichel,fenichel1,fenichel2,wiggins,wiggins1,wiggins2,eldering}. Then, the NHIM of the system and its stable and unstable invariant manifolds persist when the symmetries of the system are broken and the dynamics becomes chaotic. 

The article is organized as follows: In section 2, we show the Hamiltonian function of the system in the Boyer-Lindquist coordinates and the integrals of motion of the system. In section 3, we construct the NHIM for the non-rotating black hole, $a=0$,  using the effective potential energy obtained from the equation for the motion of the radial coordinate. Also, we show plots of the stable and unstable manifolds of the NHIM using a visualization technique based on the arclength of the trajectories in phase space known as Lagrangian descriptors \cite{wiggins5,wiggins6,gonzalez3,mancho,lopesinos}.  In section 4, we break the spatial spherical symmetry including the rotation of the black hole, $a>0$, and visualize the changes in the invariant manifolds. Finally, we present the conclusions and final remarks in the last section.

\newpage

\section{ Hamiltonian Function and Integrals of Motion}

A convenient set of coordinates to express the metric of the systems is the Boyer-Lindquist coordinates $r,\theta,\phi,t$ \cite{boyer}. These coordinates simplify the expression of the Kerr metric and facilitate the analysis of the phase space of the system \cite{wiltshire,carter,bizyaev,levin1,levin2, levin}. In these coordinates, the interval has only one mixed term.  The Kerr metric with natural units $G, c =1$, in the  Boyer-Lindquist coordinates is given by
\begin{eqnarray}
ds^2 =-m^2 d\tau^2=
&-&\left (1-\frac{2Mr}{\Sigma}\right )dt^2
-\frac{4Mar\sin^2\theta}{\Sigma}dtd\phi \\
&+&\frac{\Sigma}{\Delta}dr^2+\Sigma d\theta^2 \nonumber 
+\sin^2\theta\left
(r^2+a^2+\frac{2Ma^2r\sin^2\theta}{\Sigma}\right ) d\phi^2 \;, 
\label{metric}
\end{eqnarray}

\noindent where  $\tau$ is the proper time of the particle, $m$ and $M$, are the mass of the particle and the black hole respectively, $a$ is the spin parameter of the black hole and the functions $\Sigma$ and  $\Delta$ are defined as

\begin{eqnarray}
\Sigma(r,\theta) &= & r^2+a^2\cos^2\theta \;, \nonumber \\
\Delta(r) &= & r^2-2Mr+a^2  \; .  
\end{eqnarray}

\noindent In order to simplify the notation, we use the convention $M=1$. Using the metric we can calculate the Hamiltonian function of a particle around a rotating black hole. In particular, we use the Hamiltonian function constructed in the reference \cite{levin}. The Hamiltonian function of the systems in Boyer-Lindquist coordinates is given by

\begin{equation}
H(r,\theta,\phi,t,p_r,p_{\theta},p_{\phi},p_t) =
\frac{\Delta}{2\Sigma}p_r^2+\frac{1}{2\Sigma}p_\theta^2
-\frac{R + \Delta\Theta}{2\Delta \Sigma} -\frac{1}{2} \;,
\label{niceham}
\end{equation}

where the functions $\Theta$, $R$ are defined as

\begin{eqnarray}
\Theta(\theta) &=& Q - \cos^2\theta \; \left
\{a^2(1- E^2)+\frac{L^2}{\sin^2\theta}\right \} \;, \nonumber \\
R(r) &=& P^2-\Delta \left\{  r^2+(L-aE)^2 + Q \right\} \;,  \nonumber \\
P(r) &=& E(r^2+a^2)-aL \;.
 \quad 
\label{dimpots}
\end{eqnarray}

The corresponding Hamilton's equations of motion parametrized by the proper time $\tau$ of the particle are

\begin{eqnarray}
\dot{r}&=& \frac{\Delta}{\Sigma} p_r, \nonumber  \\
\dot{p}_r&=& \frac{\partial}{\partial r} \left(\frac{\Delta}{2\Sigma}\right) p^2_r - \frac{\partial}{\partial r} \left( \frac{1}{2\Sigma} \right)  p^2_{\theta} +  \frac{\partial}{\partial r} \left( \frac{R+\Delta \Theta}{2\Delta\Sigma} \right), \nonumber\\
\dot{\theta}&=& \frac{1}{\Sigma} p_{\theta},\nonumber\\
\dot{p}_{\theta}&=& \frac{\partial}{\partial \theta} \left(\frac{\Delta}{2\Sigma} \right) p^2_r - \frac{\partial}{\partial \theta}   \left( \frac{1}{2\Sigma} \right) p^2_{\theta} +  \frac{\partial}{\partial \theta} \left( \frac{R+\Delta \Theta}{2\Delta\Sigma} \right), \nonumber\\
\dot{\phi}&=& - \frac{1}{2\Delta\Sigma} \frac{\partial}{\partial E} \left( R+\Delta\Theta \right) , \nonumber\\
\dot{p}_{\phi}&=& 0 , \nonumber\\
\dot{t}&=&  - \frac{1}{2\Delta\Sigma} \frac{\partial}{\partial L} \left( R+\Delta\Theta \right), \nonumber\\
\dot{p}_{t}&=& 0. \label{eq:hamilton}
\end{eqnarray}

\noindent This Hamiltonian system is integrable, and the equations of motion have 4 independent integrals:  the value of the Hamiltonian $H$, the energy of the particle $E=-p_t$ , the angular momentum of the particle $L= p_\phi$, and the Carter integral $Q$. The first 3 integrals are obtained immediately from the structure of the Hamiltonian, however, the last one, $Q$ was obtained by Carter in \cite{carter} solving the Hamilton-Jacobi equation using the technique of separation of variables with another set of canonical coordinates \cite{rana}. From that approach is possible to obtain the first-order geodesic equations in a natural way. Those ordinary differential equations in the Boyer-Lindquist coordinates are given by

\begin{eqnarray}
m\Sigma \dot{r}       &=& \pm \sqrt{R}  \; \nonumber, \\
m\Sigma \dot{\theta}  &=& \pm \sqrt{\Theta} \;, \nonumber\\
m\Sigma \dot{\phi} &=& \frac{a}{\Delta}(2rE-aL)+\frac{L}{\sin{\theta}}\;, \nonumber \\
m\Sigma \dot{t}       &=& \frac{(r^2+a^2)^2 E - 2 arL}{\Delta} -a^2E\sin{\theta}\; \label{radial}.
\end{eqnarray}

This set of differential equations is very convenient for analysing important analytical properties like bifurcations in the reduced phase space \cite{bizyaev,teo,rana,rana1}. However, for the numerical calculations in the present article, we are going to use Hamilton's equations of motion Eq. (\ref{eq:hamilton}), like in works \cite{levin1,levin2,levin}. We use the equations of motion obtained directly from the Hamiltonian function Eq.  (\ref{niceham}) because we want to understand properties relevant to non-integrable cases, where all the equations of motion are coupled. Also, from the numerical point of view, the ODE system Eq.  (\ref{eq:hamilton}) is easier to code because there are no change signs related to the square roots like in Eq.   (\ref{radial}).

Also, we can construct a reduced 6-dimensional phase space to describe the dynamics of the system using the global time $t$ to parametrize the solutions of the systems of differential equations. To construct the system of differential equations for that reduced phase space, we only need to use the differential equation $\dot{t}=\frac{dt}{d\tau}$ and the chain rule $\frac{d}{dt}= \frac{d \tau}{d t} \frac{d}{d\tau}$ to calculate the derivative with respect to $t$ of the other coordinates and momenta in the system, for more details see the reference \cite{levin2}. 

\section{NHIM and its Stable and Unstable Manifolds for the Schwarzschild Black Hole, $a=0$ }

The symmetries of the system facilitate the understanding of its multidimensional phase space. In the present case, the symmetries allow us to see the basic elements that constitute the invariant manifolds under the dynamics. Some recent examples where was possible to find the Normally Hyperbolic Invariant Manifold (NHIM) and their stable and unstable invariant manifolds using this approach are in the works \cite{wiggins1,wiggins2,gonzalez,gonzalez1,gonzalez3,firmbach,stober}. 

First, let us consider the simplest case where the black hole does not rotate, $a=0$, and then the Kerr metric is reduced to the Schwarzschild metric. In this case, the system has spatial spherical symmetry it is easy to construct and visualize the generalization of the hyperbolic periodic orbit and their invariant stable and unstable manifolds. Without loss of generality, we set $Q=0$, which corresponds to the motion on the equatorial plane and $\theta = \pi/2$. It is equivalent to any other plane due to the spherical symmetry of the Schwarzschild metric. From the geodesic equation for the radial coordinate Eq.  (\ref{radial}), we can obtain an equation for the conservation of the energy $E$. 

\begin{equation}
\frac{\dot r^2}{2}+V_{eff}(r;L)=  E_{eff} = \frac{E^2}{2} \;, 
\label{eqmr}
\end{equation}
where
\begin{equation}
   V_{eff}(r;L) = \frac{m^2}{2}
  -\frac{m}{r}+\frac{L^2}{2r^2}-\frac{L^2}{r^3} \; 
\label{veff}
\end{equation}

\noindent is the radial effective potential parametrized by the conserved component of the angular momentum $L$.  The constant $E_{eff}$ is called the effective energy. 

\begin{figure}[H]
    \centering
    \includegraphics[scale=0.6]{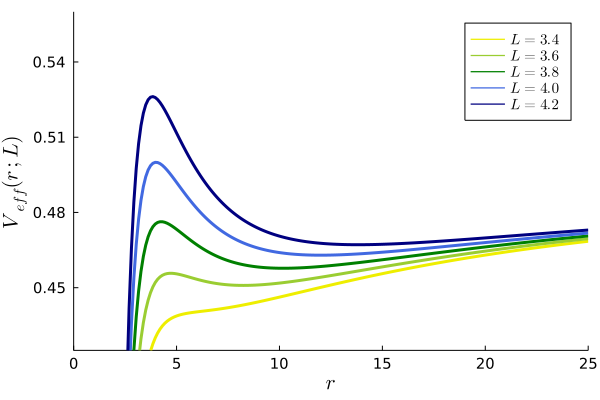}
    \caption{ Radial effective potential $V_{eff}(r;L)$ for different values of $L$. As the value of $L$ is increased the maximum value of $V_{eff}(r;L)$ grows. Associated with each maximum there is an unstable hyperbolic fixed point $X_L$ in the phase space of the one-dimensional system with energy $E=E_h$. This fixed point is located at $X_L=(r=r_L,p_r=0)$ where  $r_L$ is the critical value of $V_{eff}(r;L)$ defined by its maximum. }
    \label{fig:veff}
\end{figure}

Due to the existence of the radial effective potential $V_eff(r;L)$, let us consider the reduced phase space only for the radial degree of freedom. Associated with the maximum of the radial effective potential $V_{eff}(r;L)$ there is an unstable hyperbolic fixed point $X_L=(r=r_L,p_r=0)$. There are two special sets in the phase space that divide the reduced 2-dimensional phase space in regions with different behaviour, the stable and unstable manifolds of the unstable hyperbolic fixed point denoted by $W^{s/u}(X_h)$. The stable (unstable) manifold of the hyperbolic fixed point is the set trajectories that converge (diverge) when the time is increased, for this 1-degree-of-freedom system is just a single trajectory. 

When we include the angular coordinate $\phi$ in the analysis of the dynamics, we have a 2-degree-of-freedom system. The unstable hyperbolic fixed point $X_L$ corresponds to a circular unstable hyperbolic periodic orbit $\gamma_L$ with radius $r_L$ in the configuration space. The unstable hyperbolic periodic orbit $\gamma_L$ also has stable and unstable manifolds $W^{s/u}(\gamma_L)$. Those 2-dimensional invariant manifolds are formed by the union of all the trajectories that converge (diverge) to $\gamma_L$ when the time to infinity. Also by construction,  $W^{s/u}(\gamma_L)$ divides the constant energy manifold of the 2-degree-of-freedom system. 

Considering the spatial spherical symmetry of the system, we construct the generalization of this unstable hyperbolic periodic orbit in the constant energy manifold, a Normally Hyperbolic Invariant Manifold (NHIM) denoted as $M_{E,a=0}$. Let us take the union of all the unstable circular periodic orbits with the energy $E$, associated with the maximum of the effective radial potential $V_{eff}(r;L)$, for all the possible angular directions $\theta,\phi$ on the sphere $S^2$ in the configuration space give us

\begin{eqnarray}
   M_{E,a=0} &=& \bigcup_{\theta,\phi} \gamma_L = S^2 \bigtimes \gamma_L \;.
\end{eqnarray}

\noindent By construction, the topology of the NHIM $M_{E,a=0}$ is a 3-dimensional sphere $S^3$. If we consider the degree of freedom corresponding to the time $t$ in the construction of the NHIM, the result is a 4-dimensional unbounded NHIM given by $ M_{E,a=0} \bigtimes {\rm I\!R} $.

Analogously, we construct the stable and unstable manifolds of the NHIM $M_{E,a=0}$.  Again, considering the spherical spacial symmetry of the system and taking the union of all the stable or unstable manifolds of $W^{s/u}(\gamma_L)$, we obtain the invariant manifolds of the NHIM $M_{E,a=0}$ given by

\begin{eqnarray}
   W^{s/u}(M_{E,a=0}) &=& \bigcup_{\theta,\phi} W^{s/u}(\gamma_L) \;.
\end{eqnarray}

\noindent The topology of $W^{s/u}(M_{E,a=0})$ is a 4-dimensional spherical cylinder $S^3 \times {\rm I\!R}$ by construction.  This spherical cylinder divides the 5-dimensional constant energy manifold in the reduced phase space that considers only the spatial degrees of freedom. Also, if we consider the degree of freedom corresponding to $t$ in the construction of the invariant stable and unstable manifolds of the NHIM, the result is a 5-dimensional invariant manifold given by $ W^{s/u}(M_{E,a=0}) \bigtimes {\rm I\!R} $. 

In order to visualize the invariant manifolds in multidimensional phase space, we calculate the scalar field of the arclength of the trajectories.  The different behaviours of the trajectories generate differences in the scalar field of arclength giving us information on the invariant manifolds that intersect the set of initial conditions where we calculate the scalar field. This scalar field is a type of Lagrangian descriptor. Some recent examples where the Lagrangian descriptors have been used to analyze the phase space of multidimensional systems are in \cite{wiggins5,gonzalez3,gonzalez2,gonzalez4}. The basic definition of the Lagrangian descriptor based on arclength and details about the numerical calculations for this work are in Appendix A, also more illustrative examples where this technique has been used to study the phase space of different chaotic systems are in the reference therein.  

Let us consider some relevant values of $L$ where the phase space changes qualitatively. In this case, the changes in the phase space are related to changes in the geometry of the radial effective potential energy $V_{eff}(r;L)$. In Fig. \ref{fig:veff}, we can see two important changes in the geometry of the radial effective potential energy $V_{eff}(r;L)$ as the value of $L$ is increased.  

\begin{figure}[H]

    \centering
    \begin{tabular}{c c}
     \subf{\includegraphics[scale=0.365]{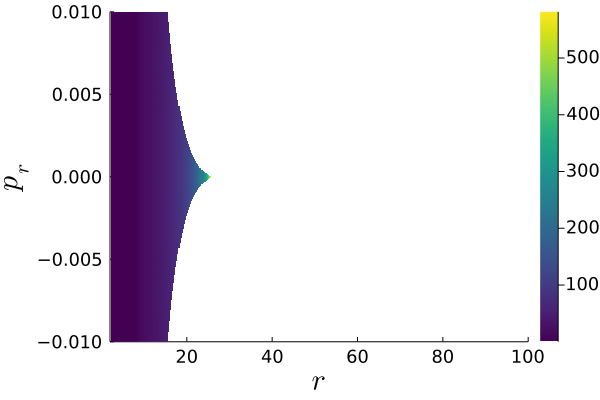}}
     {a) $L=3.4641 < L_{sc}$ }
     &
     \subf{\includegraphics[scale=0.365]{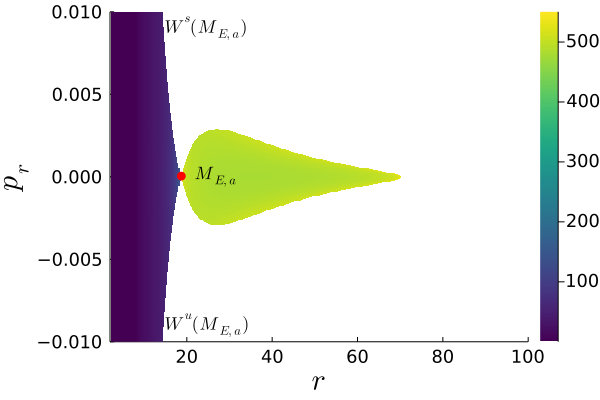}}
     {b) $L=3.6  > L_{sc}$ , $E=E_h$ }
     \\
     \subf{\includegraphics[scale=0.365]{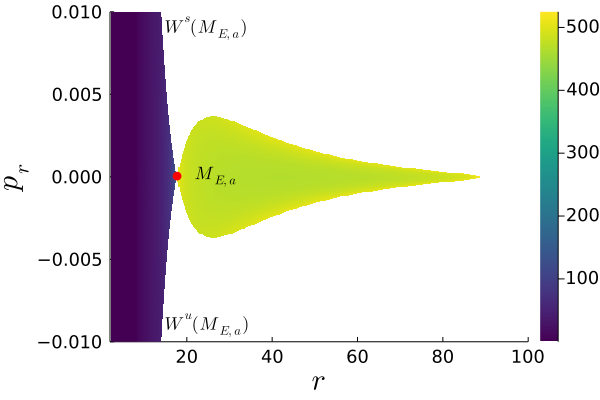}}
     {c) $L=3.65$, $E=E_h$}
     &    
     \subf{\includegraphics[scale=0.365]{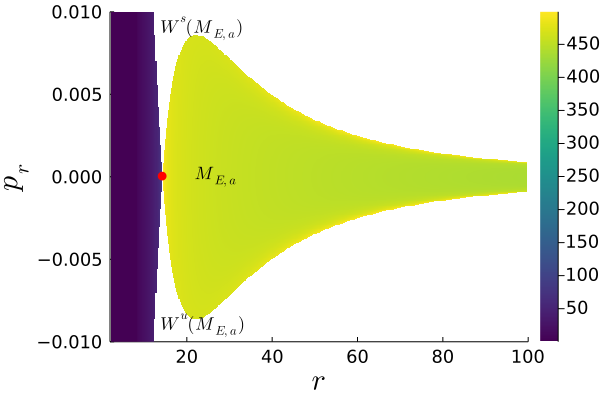}}
     {d) $L=4.0$, $E=E_h$}
    \end{tabular}
     
    \caption{ Lagrangian descriptor evaluated on the plane $r$--$p_r$ for different values of $E$, $L$, $p_\theta=0$, and $a=0$. The energy $E=E_h$ corresponds to the energy of the hyperbolic fixed point associated with the maximum of the radial effective potential $V_{eff}(r;L)$. The red point is the intersection of the initial conditions with the NHIM $M_{E,a=0}$. The internal branches of the stable and unstable manifolds $W^{s/u}(M_{E,a=0})$ divide the internal region in dark blue from the forbidden region in the domain in white. The external branches of $W^{s/u}(M_{E,a=0})$ divide the external area in green--yellow from the forbidden region.  }
    \label{LD_L_E}
    
\end{figure}

The first important change in the radial effective potential  $V_{eff}(r;L)$ is the collision between its maximum and minimum at $L=L_{sc}=\sqrt{12}$. This collision generates a saddle-centre bifurcation between the associated periodic orbits in the phase space. To visualize those changes in the phase space, we calculate the Lagrangian descriptor for initial conditions in the canonical plane $r$--$p_r$ and two constant values of $L$, one slightly smaller than the critical value and the other slightly bigger, see Fig. \ref{LD_L_E}  panels a) and b). We can see how the phase space grows abruptly and a bounded region appears. In Fig. \ref{LD_L_E}  b), the boundaries of phase space are the intersection of invariant manifolds $W^{s/u}(\gamma_L)$ with the set of initial conditions and the intersection point marked as the red point is the intersection with the unstable hyperbolic periodic orbit $\gamma_L$ that is contained in the NHIM $M_{E,a=0}$. 

The second important change in  $V_{eff}(r;L)$ occurs when its maximum reach the asymptotic value $1/2$. For example if $L=4$, the trajectories can escape to infinity and the phase space has a transition from bounded to unbounded. Analogously, we calculate the Lagrangian descriptor for initial conditions in the canonical plane $r$--$p_r$ and two constant values of $L$, one value slightly smaller than the critical value and other chosen value slightly bigger, see Fig. \ref{LD_L_E} panels c) and d). In Fig. \ref{LD_L_E}  d), we can see the external branches of invariant manifolds $W^{s/u}(\gamma_L)$ going to infinity.

\begin{figure}[H]

    \centering
    \begin{tabular}{c c}
     \subf{\includegraphics[scale=0.365]{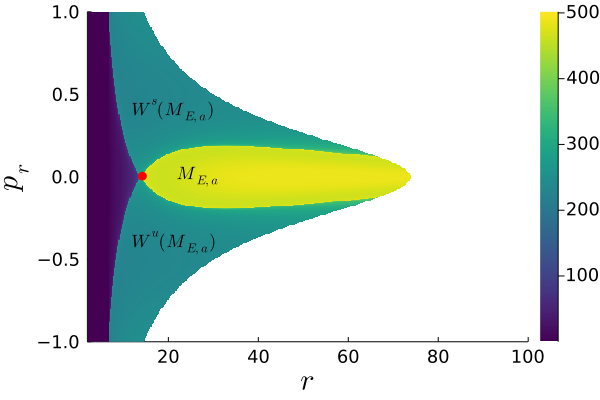}}
     {a) $E_{eff}=0.44$}
     &
     \subf{\includegraphics[scale=0.365]{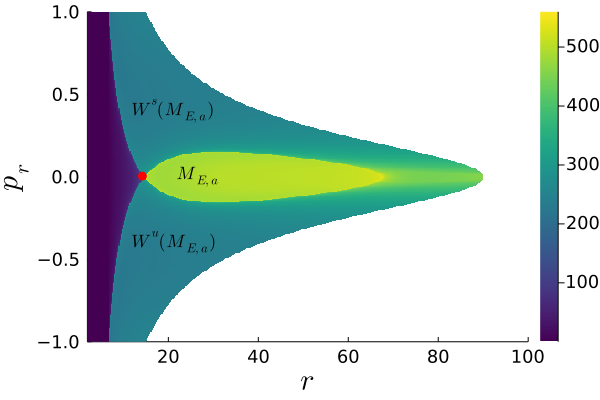}}
     {b) $E_{eff}=0.45$}
     \\
     \subf{\includegraphics[scale=0.365]{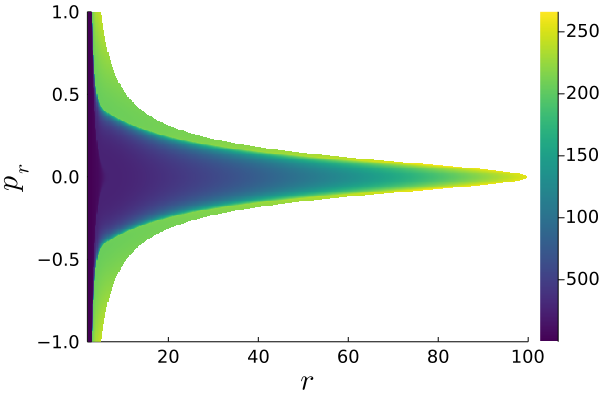}}
     {c) $E_{eff}=0.49$}
     &    
     \subf{\includegraphics[scale=0.365]{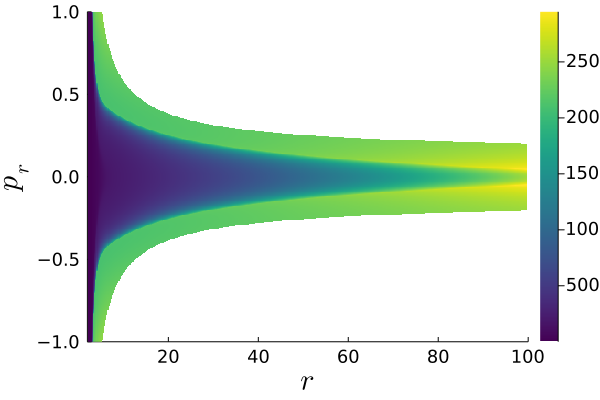}}
     {d) $E_{eff}=0.51$}
    \end{tabular}
     
    \caption{ Lagrangian Descriptor evaluated on the plane $r$--$p_r$ for different values of $E$ and $a=0$. For panels a) and b),  the red dot is the intersection of the initial conditions with the NHIM $M_{E,a=0}$. The internal branches of the stable and unstable manifolds $W^{s/u}(M_{E,a=0})$ divide the internal region in dark blue from the forbidden region in white. The external branches of $W^{s/u}$ divide the external area in green--yellow from the forbidden region. For panels c) and d) there is not a NHIM.}
    \label{fig:ld_E=cst_a=0}.
    
\end{figure}

Now, we consider the Lagrangian descriptor evaluated on initial conditions in the plane $r$---$p_r$ with the initial component of the momentum $p_\theta=0$ and the value of the energy $E$ also constant.  Let us remark that for these calculations the value of $L$ is determined by the other initial conditions and its value changes from point to point in the next plots. The Fig. \ref{fig:ld_E=cst_a=0} shows the results for 4 different values of $E$ or equivalently the effective energy $E_{eff}$. 

For the value of the effective energy $E_{eff} = 0.44$,  see Fig. \ref{fig:ld_E=cst_a=0} a), the Lagrangian descriptor plot has 4 regions with different dynamics: the yellow region corresponds to bounded trajectories, the dark--blue region corresponds to trajectories that cross the event horizon and go inside the black hole, and the 2 green regions correspond to unbounded trajectories. Let us remark that the NHIM $M_{E,a=0}$ is the intersection of the stable and unstable manifolds $W^{s/u}(M_{E,a=0})$ and the intersection of  $M_{E,a=0}$ with the set of initial conditions is the red point. The internal branches of the stable and unstable manifolds $W^{s/u}(M_{E,a=0})$ are the lines that divide the dark--blue and the green regions and external branches of $W^{s/u}(M_{E,a=0})$ are the boundaries of the trapped yellow region. In this case, the trapped region reaches the boundary of the domain of the Lagrangian descriptor plot. 

Increasing a little the value of the effective energy to $E_{eff} = 0.45$, see Fig. \ref{fig:ld_E=cst_a=0} b),  we have  3 regions on the Lagrangian descriptor plot. The bounded region does not reach the boundary of the domain of the plot. The intersection of the external branches of invariant manifolds $W^{s/u}(M_{E,a=0})$ with the set of initial conditions form a closed separatrix curve that defines the boundary of the trapped region. There is only one region where the trajectories are unbounded.

For the effective energy $E_{eff}=0.49$, Fig. \ref{fig:ld_E=cst_a=0} c) shows that the trapped region inside the separatrix formed by the external branches of $W^{s/u}(M_{E,a=0})$ has disappeared and the region that goes to the event horizon grows, however the domain of the Lagrangian descriptor plot still bounded. In this case, the yellow--green regions correspond to regions with a large value of final $r$.

For $E_{eff}>1/2$ the phase space becomes unbounded. We see in Fig. \ref{fig:ld_E=cst_a=0} d) for $E_{eff}=0.51$ that the domain of the Lagrangian descriptor plot is unbounded, but the phase space has a similar structure that the panel c).  

\section{NHIM and its Invariant Stable and Unstable Manifolds for Kerr Black Hole, {\bf $a>0$ } }

For the rotating black hole, its angular momentum is non-zero, i.e. $a>0$,  the system loses its spatial spherical symmetry, however, it is still an integrable one \cite{carter}. A natural question is: What happens to the NHIM $M_{E,a=0}$ and their stable and unstable manifolds $W^{s/u}(M_{E,a=0})$? In general, like in the case of the loss of the spherical symmetry for the existence of the black hole \cite{penrose}, the NHIMs and their stable and unstable manifolds are robust under perturbations of the vector field that define the system of ordinary differential equations \cite{fenichel,wiggins}. This fundamental geometrical property has been essential to understanding the dynamics in the multidimensional phase space for different systems where symmetry is broken, see for example \cite{gonzalez,gonzalez1,gonzalez3,drotos,drotos1}.

Let us denote the NHIM for $a>0$ as $M_{E,a}$ and its stable and unstable invariant manifolds as $W^{s/u}(M_{E,a})$. Recent works based on the integrability of the system and a change of variables \cite{rana,teo,stein} show analytical expressions for the trajectories contained in the NHIM $M_{E,a}$ and their stable and unstable manifolds $W^{s/u}(M_{E,a})$  for $a\in[0,1]$.

Moreover, we can understand the geometrical origin of the existence of the NHIM and its stable and unstable manifolds analogously than in the non-rotating case, $a=0$,  considering again the Eq.  (\ref{radial}) for the radial motion and the dynamical systems theory. In this case, the equation for the radial motion can be written as

\begin{equation}
\frac{m^2\Sigma^2}{r^4}\frac{\dot r^2}{2}+V_{eff}(r;a,E,L,Q) =  E_{eff} = \frac{E^2}{2} \;,
\label{eqmra}
\end{equation}
where
\begin{equation}
   V_{eff}(r;a,E,L,Q) = \frac{m^2}{2}
  -\frac{m}{r}+\frac{L^2-a^2(E^2-m^2) + Q }{2r^2} - \frac{(L-aE)^2 + Q}{r^3} + \frac{a^2Q}{2r^4} \;. 
\label{veffa}
\end{equation}

\noindent On the RHS of Eq.  (\ref{eqmra}) we have the radial kinetic energy and the generalized radial effective potential energy $V_{eff}(r;a,E,L,Q)$. On the LHS of Eq.  (\ref{eqmra}) we have a constant interpreted as the effective energy.

Let us notice that for $a=0$ and $Q=0$, the radial effective potential Eq.  (\ref{veffa}) and the radial equation of motion Eq.  (\ref{eqmra}) are reduced to the Eq.  (\ref{veff}) and Eq.  (\ref{eqmr}) in the previous section. Also, we obtain the same functional form for the radial effective potential with the change $ L^2 + Q \to L^2 $ for the case $a=0$ and $Q \neq 0$. That represents the motion outside the equatorial plane in the non-rotational case. For detailed information about the bifurcation diagrams for all the parameters see the references \cite{bizyaev,teo}.

\begin{figure}[H]
    \centering
    \includegraphics[scale=0.6]{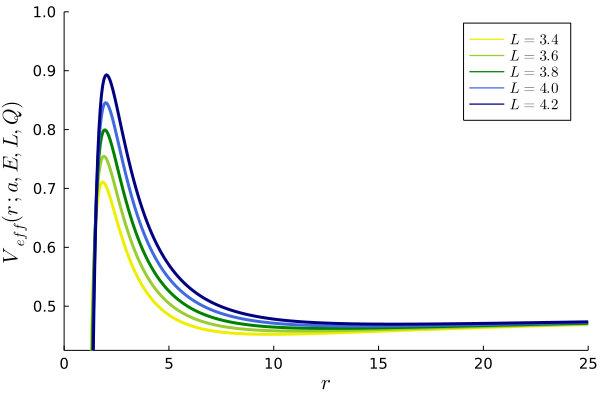}
    \caption{Radial effective potential $V_{eff}(r;a,E,L,Q)$ for $a=1$, $E=0.9428$, $Q=0$, and different values of $L$. Associated with each maximum there is an unstable hyperbolic fixed point in the phase space of the 1-dimensional system. This fixed point is located at $(r,p_r)=(r_{(a,E,L,Q)},0)$ where  $r_{(a,E,L,Q)}$ is the local maximal value of $V_{eff}(r;a,E,L,Q)$. }
    \label{fig:veff2}
\end{figure}

Let us remark that we can apply the theorem of persistence of NHIMs and their invariant manifolds to $M_{E,a}$  and  $W^{s/u}(M_{E,a})$  for a perturbed Hamiltonian generated by a perturbed metric.  Typically, the system becomes non-integrable and its dynamics is chaotic under a generic perturbation. Then the stable and stable manifolds  $W^{s/u}(M_{E,a})$  intersect transversally an infinite number of times and form a tangle in the phase space. This tangle is the multidimensional generalisation of the Smale horseshoes for the Poincare map of the 2-degree-of-freedom system, some examples are found in the references \cite {gonzalez,drotos,zapfe,gonzalez4}. 

\begin{figure}[H]

    \centering
    \begin{tabular}{c c}
     \subf{\includegraphics[scale=0.365]{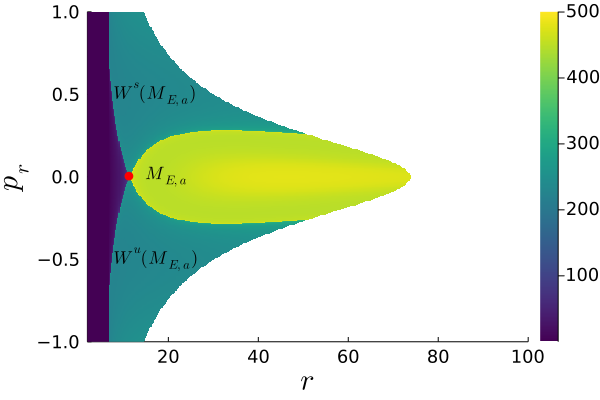}}
     {$a=0.1$}
     &
     \subf{\includegraphics[scale=0.365]{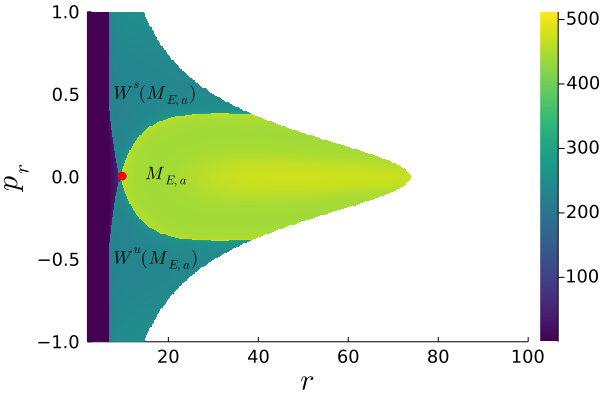}}
     {$a=0.2$}
     \\
     \subf{\includegraphics[scale=0.365]{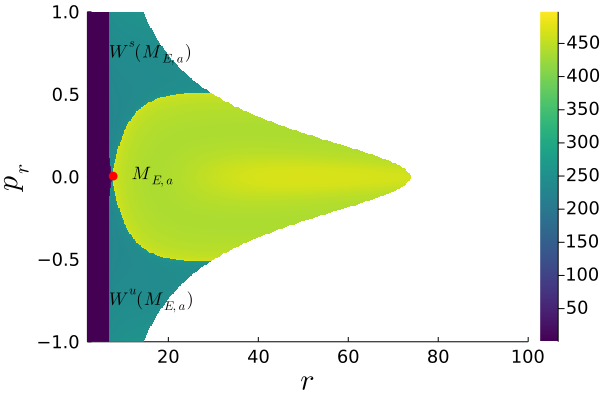}}
     {$a=0.3$}
     &    
     \subf{\includegraphics[scale=0.365]{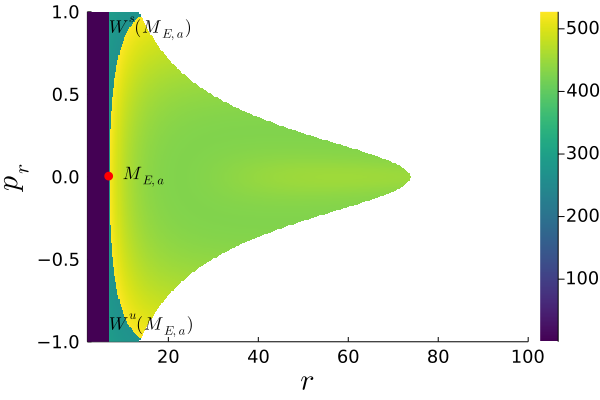}}
     {$a=1.0$}
    \end{tabular}
    \caption{ Lagrangian Descriptor evaluated on the plane $r$--$p_r$ for $E_{eff}=0.44$, different values of $a$. }
    \label{fig:ld_Eeff=044}
\end{figure}

\begin{figure}[H]
    \centering
    \begin{tabular}{c c}
     \subf{\includegraphics[scale=0.365]{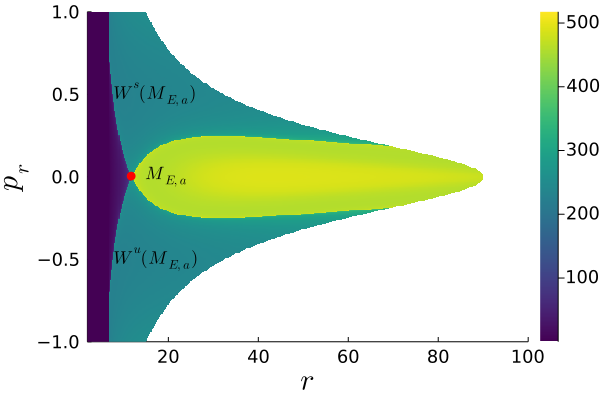}}
     {$a=0.1$}
     &
     \subf{\includegraphics[scale=0.365]{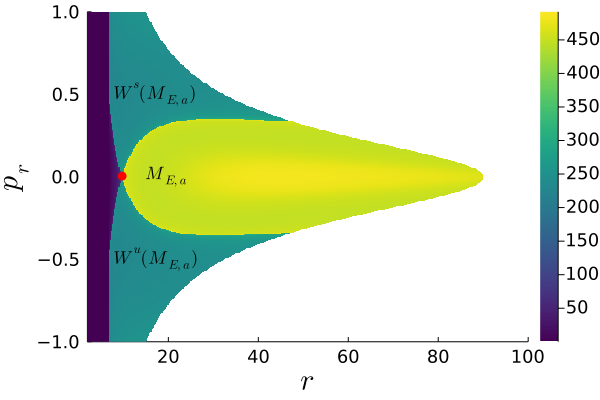}}
     {$a=0.2$}
     \\
     \subf{\includegraphics[scale=0.365]{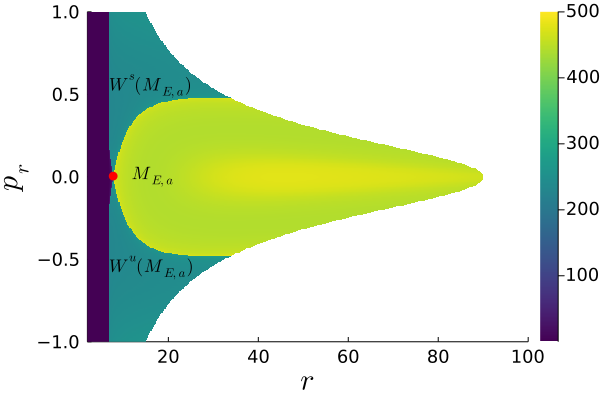}}
     {$a=0.3$}
     &    
     \subf{\includegraphics[scale=0.365]{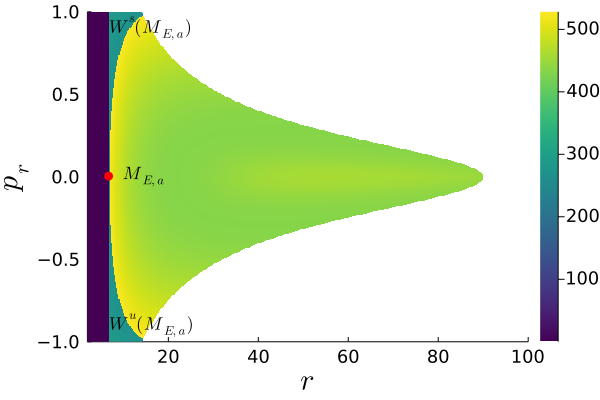}}
     {$a=1.0$}
    \end{tabular}
     
    \caption{ Lagrangian Descriptor evaluated on the plane $r$--$p_r$ for $E_{eff}=0.45$, different values of $a$. }
    \label{fig:ld_Eeff=045}
\end{figure}

\begin{figure}[H]

    \centering
    \begin{tabular}{c c}
     \subf{\includegraphics[scale=0.365]{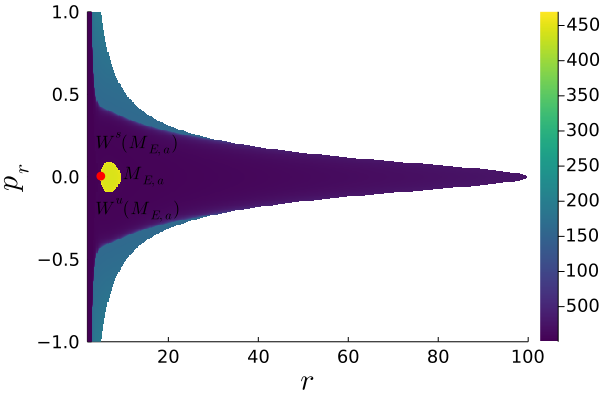}}
     {$a=0.1$}
     &
     \subf{\includegraphics[scale=0.365]{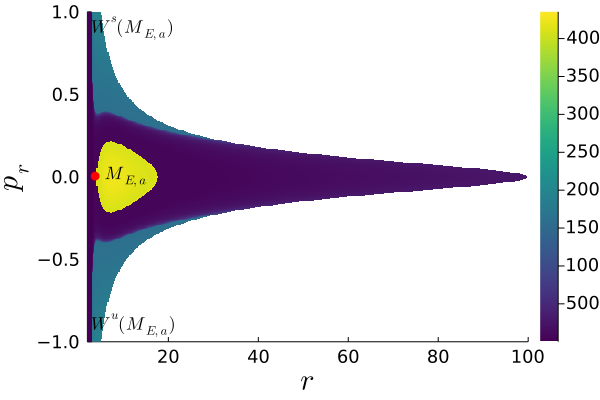}}
     {$a=0.2$}
     \\
     \subf{\includegraphics[scale=0.365]{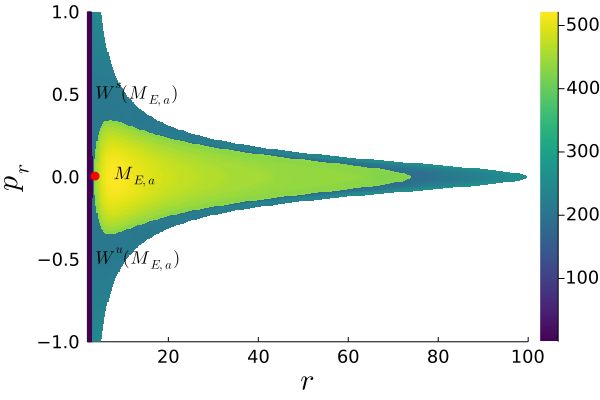}}
     {$a=0.3$}
     &    
     \subf{\includegraphics[scale=0.365]{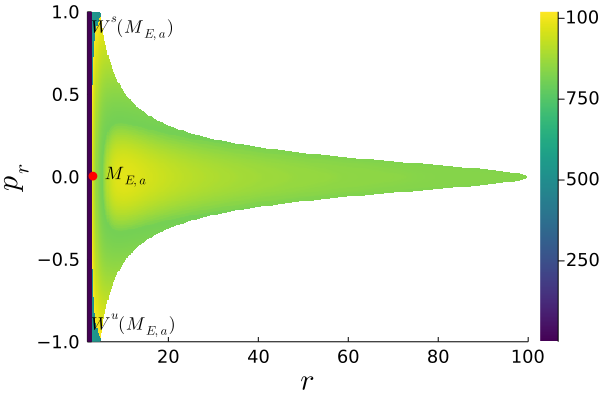}}
     {$a=1.0$}
    \end{tabular}
         
    \caption{ Lagrangian Descriptor evaluated on the plane $r$--$p_r$ for $E_{eff}=0.49$, different values of $a$.  }
    \label{fig:ld_Eeff=049}
\end{figure}

\begin{figure}[H]
    \centering
    \begin{tabular}{c c}
     \subf{\includegraphics[scale=0.365]{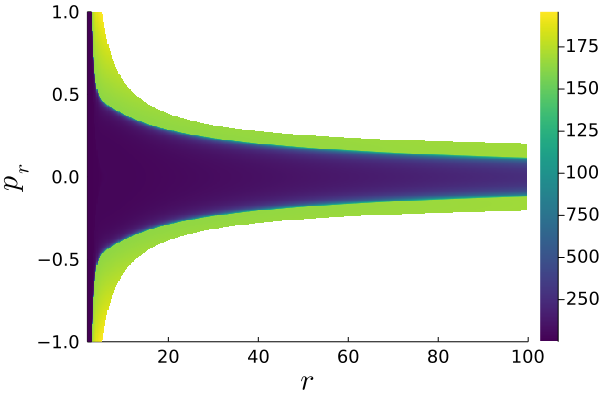}}
     {$a=0.1$}
     &
     \subf{\includegraphics[scale=0.365]{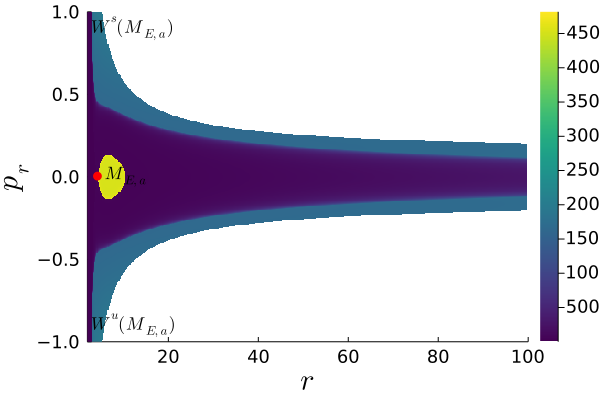}}
     {$a=0.2$}
     \\
     \subf{\includegraphics[scale=0.365]{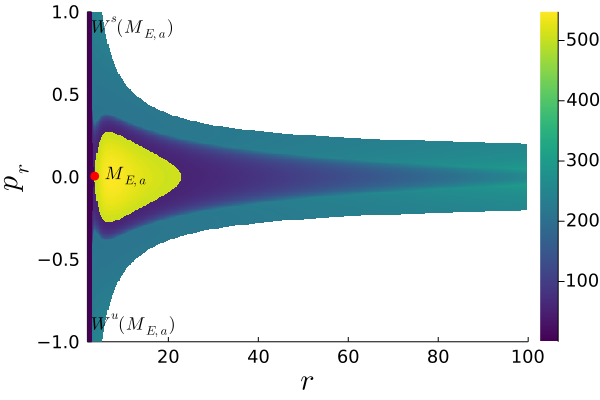}}
     {$a=0.3$}
     &    
     \subf{\includegraphics[scale=0.365]{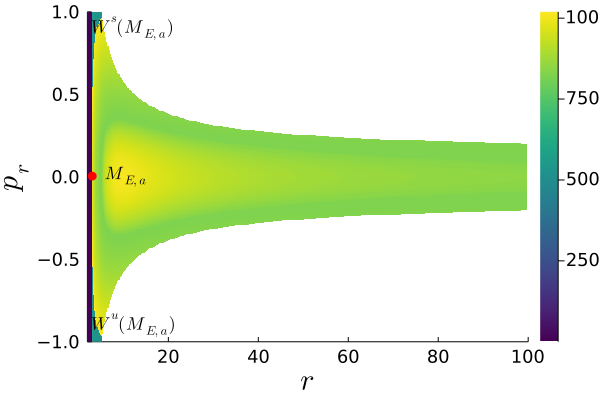}}
     {$a=1.0$}
    \end{tabular}
    \caption{ Lagrangian Descriptor evaluated on the plane $r$--$p_r$ for $E_{eff}=0.51$, different values of $a$. }
    \label{fig:ld_Eeff=051}
\end{figure}

The Figs. \ref{fig:ld_Eeff=044},\ref{fig:ld_Eeff=045},\ref{fig:ld_Eeff=049},\ref{fig:ld_Eeff=051} show typical changes in the phase space for different values of the effective energy of the particle $E_{eff}$ and different values of the rotation parameter $a$ of the black hole. In general, we can appreciate how the trapped region grows when the rotation parameter of the black hole $a$ is increased and, in contrast, how the size of the trapped region decreases when the energy of the particle grows. For the effective energy $E_{eff}=0.44$, see Fig. \ref{fig:ld_Eeff=044}, the domain of the Lagrangian descriptor plot is bounded and the separatrix is defined by the invariant manifolds $W^{s/u}(M_{E,a})$. For $E_{eff}=0.45$, see Fig. \ref{fig:ld_Eeff=045}, the domains of the Lagrangian descriptor plots and the trapped regions are bigger than in the previous effective energy, however, the phase space structure is similar. For $E_{eff}=0.49$, see Fig. \ref{fig:ld_Eeff=049},  the trapped regions do not reach the maximum value of $r$ in the domain of the Lagrangian descriptor plots for moderate values of $a$. Finally, for the value $E_{eff}=0.51$, the domain of the Lagrangian descriptor becomes unbounded, see Fig. \ref{fig:ld_Eeff=051}. However, the structure of the trapped region looks similar to the previous case for small enough values of $a$.

\newpage

\section{Conclusions and Remarks}

In this work, we have studied the phase space of a particle moving around a rotating Kerr black hole using the  Boyer-Lindquist coordinates. The dynamics of this system is integrable and has 4 conserved quantities:  the value of the Hamiltonian $H$, the energy of the particle $E=-p_t$ , the angular momentum $L= p_\phi$, and the Carter integral $Q$. From the radial first integral equation is possible to find a radial effective potential $V_{eff}(r;a,E,L,Q)$  parametrized by the rotation parameter of the black hole $a$ and the conserved quantities of the system. Using the radial effective potential, we have constructed a family of unstable periodic orbits associated with its maximum value for a fixed value of $E$.  The union of all those unstable hyperbolic periodic orbits form a Normally Hyperbolic Invariant Manifold (NHIM)  $M_{E,a}$ that has a central role in the dynamics of the system. 

The stable and unstable manifolds $W^{s/u}(M_{E,a})$  have codimension one relative to the constant energy manifold and divide it into regions with different behaviour, see Figs. \ref{fig:ld_E=cst_a=0},\ref{fig:ld_Eeff=044},\ref{fig:ld_Eeff=045},\ref{fig:ld_Eeff=049}, \ref{fig:ld_Eeff=051}. Due to the integrability of the system, the external branches of the stable and unstable manifolds $W^{s/u}(M_{E,a})$ coincide and form a separatrix that encircles the KAM tori associated with the minimum of the radial effective potential. The internal branches of $W^{s/u}(M_{E,a})$ form an impenetrable barrier in the phase space that defines the set of trajectories that can cross the event horizon. In one way, the phase space of this system is similar to other multidimensional scattering systems \cite{gonzalez3,gonzalez,drotos,zapfe}, but in this case the open branches of $W^{s/u}(M_{E,a})$ go to the central region instead of the asymptotic region, $r\to \infty$.  

We have analyzed the changes in the phase space of the system for different parameters using Lagrangian descriptors based on the arclength of the trajectories. The Lagrangian descriptors plots allow us to find
the NHIM $M_{E,a}$ and its stable and unstable manifolds $W^{s/u}(M_{E,a})$ easily. On those plots, we find the intersections of the $W^{s/u}(M_{E,a})$ with the set of initial conditions used to calculate the arclength of the trajectories. In this integrable system, the only one intersection between $W^{s}(M_{E,a})$ and $W^{u}(M_{E,a})$ is the NHIM $M_{E,a}$. In particular, the Lagrangian descriptor plots show that when the rotation of the black hole $a$ grows the trapped region also grows. Therefore, when the angular momentum of the black hole grows more it drags more of the particles around it.

In the phase space, there is a NHIM constructed with prograde orbits, $L>0$, and another one constructed with retrograde orbits, $L<0$. An important difference between the cases with $a>0$ and $a=0$ is that the NHIM constructed from the prograde orbits and the NHIM constructed from the retrograde orbits have different spatial projections because the radial effective potential energy $V_{eff}(r;a,E,L,Q)$ is not symmetric function respect the change $L \to -L$.

All the results for bounded NHIMs in the reduced phase space defined by the spatial degrees of freedom $r,\theta,\phi$ can be generalized to the case when the time $t$ is included as a degree of freedom.  In this case, is necessary to consider the theory for unbounded NHIMs and their invariant stable and unstable manifolds developed in \cite{eldering}.

In future works, we are going to study the phase space of a perturbed Kerr black hole like in \cite{lukes}. When the metric is perturbed, the symmetries are broken and the dynamics of the system becomes chaotic. However, the NHIMs and their stable and unstable invariant manifolds are robust under perturbations. This fundamental property allows us to analyse the perturbed chaotic systems. For a non-integrable perturbation, the stable and unstable manifolds of NHIM intersect transversely and a form multidimensional rich structure analogous to the Smale horseshoes in the Poincare map of the 2-degree-of-freedom systems. The stable and unstable manifolds form a system of tubes that determine the transport in the phase space. Also, the internal dynamics of the NHIM becomes chaotic and give us information about the bifurcations of the NHIM and its stable and unstable invariant manifolds.

Also, the NHIMs $M_{E,a}$ define a dividing surface in the phase space like in Wigner's Transition State Theory for chemical reaction dynamics in phase space \cite{wigner, evans, eyring, waalkens,katsanikas,katsanikas1,katsanikas2,katsanikas3,reiff}. Using this surface in the phase space is possible to calculate the total flux of trajectories that cross the saddle in potential energy and can reach the external event horizon of the black hole.

\section{Acknowledgments}

This research was funded by DGAPA UNAM grant number AG--101122, CONAHCyT CF--2023--G--763, and CONAHCyT fronteras grant number 425854. The author thanks
the Virtual Institute of Physics at New York (VIP--NY), and Centro Internacional de Ciencias AC--UNAM for their facilities during the early stages of this work. 

\newpage

\section{Appendix A: Lagrangian descriptors a tool to visualize the phase space}

The Lagrangian descriptors are scalar fields useful to visualize invariant objects in the multidimensional phase space. The main idea behind the method is based on the distinctive behaviour of the trajectories in different regions of the phase space \cite{wiggins5,wiggins6,gonzalez3,mancho,lopesinos}. Some recent illustrative examples using this phase space visualisation technique are in the references \cite{crossley,hillebrand,zimper}.  For instance, let us consider two trajectories with close initial conditions. One trajectory is contained in a stable manifold of an unstable periodic orbit $W^s(\gamma)$ and the other trajectory has nearby initial condition, but it is not contained in $W^s(\gamma)$. The trajectory that belongs to $W^s(\gamma)$ converges to $\gamma$, meanwhile the other trajectory is close to the first one only for a finite interval of time.  This different behaviour generates a significant difference in the value of their arclength that is easy to appreciate on a plot.

 The Lagrangian descriptor evaluated on the point  $x_0$ in the phase space is defined as  

\begin{equation}
    LD[\textbf{x}_0;\tau_+,\tau_-,t_0] = \int^{t_{0}+\tau_+}_{t_0} \sqrt{  \sum_j  (\dot{x}_j(t))^2  }   dt \;  + \int^{t_0}_{t_{0}-\tau_-} \sqrt{  \sum_j  (\dot{x}_j(t))^2  }   dt \;,
\end{equation}

 \noindent where $\textbf{x}(t)$ is the solution of the ODE system such that $\textbf{x}(t_0)=\textbf{x}_0$,  $\dot{\textbf{x}}_j$ are the components of the vector field that define the ODE system $\dot{\textbf{x}}=\textbf{f}(\textbf{x},t)$, and $[t_0+\tau_+,t_0-\tau_-]$ is the integration interval.
 
 In the particular example of this work, the maximal integration time possible is defined by $\tau_+=\tau_-=16000$. If the particle reaches the external event horizon of the black hole, we stop the integration to avoid numerical problems. The grid used the generate the Lagrangian descriptor pictures has $400\bigtimes400$ initial conditions. The library for the integration of the equations of motion is {\it DifferentialEquations.jl} implemented in Julia Language \cite{Julia}. The integration methods to avoid numerical problems due to the stiffness of the system are a combination of the methods {\it Tsitouras 5--4 Runge-Kutta} and {\it Rosenblock 3--2} with {\it relative tolerance $= 10^{-8}$} and {\it absolute tolerance $=10^{-12}$}.

\section{Appendix B: Stability of Periodic Orbits on the NHIM and its relation with the Effective Potential}

The NHIM $M_{E,a}$ is constructed with the periodic orbits $\gamma_L$.  
Those periodic orbits and their linear stability in the radial direction are defined by the equations

\begin{eqnarray}
\dot{r}  &=& \pm \frac{\sqrt{R}}{m\Sigma} = 0 \; ,\\
\frac{ \partial  \dot{r} }{\partial r } &>& 0   \;.
\end{eqnarray}

Substituting the definitions of $R$ and $\Sigma$, we can find that previous equations are equivalent to 

\begin{eqnarray}
R &=&0 \;,\\
\frac{d R }{d r} &=&0 \;, \\
\frac{d^2 R }{d r^2 } &>& 0 \;.   
\end{eqnarray}

In terms of the radial effective potential energy, $V_{eff}$ those conditions are related to its local maximum given by 

\begin{eqnarray}
\frac{\partial V_{eff} }{\partial r} &=&0 \;,\\
\frac{\partial^2 V_{eff} }{\partial r^2 } &<&0 \;.   
\end{eqnarray}

For more information about the parameter space and the bifurcation diagrams of the system, see the recent works \cite{bizyaev,teo,rana,rana1}.

\end{document}